# A comparison of the accuracy of saddlepoint conditional cumulative distribution function approximations

**Juan Zhang**[1,*] **and John E. Kolassa**[1,*]

*Rutgers University*

**Abstract:** Consider a model parameterized by a scalar parameter of interest and a nuisance parameter vector. Inference about the parameter of interest may be based on the signed root of the likelihood ratio statistic $R$. The standard normal approximation to the conditional distribution of $R$ typically has error of order $O(n^{-1/2})$, where $n$ is the sample size. There are several modifications for $R$, which reduce the order of error in the approximations. In this paper, we mainly investigate Barndorff-Nielsen's modified directed likelihood ratio statistic, Severini's empirical adjustment, and DiCiccio and Martin's two modifications, involving the Bayesian approach and the conditional likelihood ratio statistic. For each modification, two formats were employed to approximate the conditional cumulative distribution function; these are Barndorff-Nielson formats and the Lugannani and Rice formats. All approximations were applied to inference on the ratio of means for two independent exponential random variables. We constructed one and two-sided hypotheses tests and used the actual sizes of the tests as the measurements of accuracy to compare those approximations.

## 1. Introduction

When analyzing data arising from a model with a single unknown parameter, statisticians frequently build tests of a simple null hypothesis around the likelihood ratio statistic, since the signed square root of the likelihood ratio statistic, $R$, often has a distribution that is well-approximated by a standard normal distribution under the null hypothesis. In the presence of nuisance parameters, the statistic $R$ depends on the nuisance parameters. Practitioners often replace the nuisance parameters in the likelihood function by their maximum likelihood estimates and examine the resulting profile likelihood as a function of the parameter of interest. Denote by $n$ the sample size. The standard normal approximation to the conditional distribution of $R$ typically has error of order $O(n^{-1/2})$, and $R$ can be used to construct approximate confidence limits for the parameter of interest having coverage error of that order. In large sample settings, this approximation works well. However, in small sample situations, with 10 or 15 observations, the standard normal approximation may not be adequate. Hence, various authors developed modifications for $R$

---

*The authors were supported by Grant DMS 0505499 from the National Science Foundation. The authors are grateful for the assistance of one of the volume editors, Bill Strawderman, and an anonymous referee.

[1]Department of Statistics and Biostatistics, Hill Center, Bush Campus, Rutgers, The State University of New Jersey, 110 Frelinghuysen Rd., Piscataway, NJ 08854 USA, e-mail: janezh@stat.rutgers.edu; kolassa@stat.rutgers.edu

*AMS 2000 subject classifications:* 62E60, 41A58.

*Keywords and phrases:* modified signed likelihood ratio statistic, saddlepoint approximation, conditional cumulative distribution.

using saddlepoint approximation techniques. These modifications reduce the order of error in the standard normal approximation to the conditional distribution of $R$.

Barndorff-Nielsen [2] first proposed the modified directed signed root of the likelihood ratio statistic $R^*$. This statistic will be reviewed in the next section. The relative error in the standard normal approximation to the conditional distribution of $R^*$ is of order $O(n^{-3/2})$. Barndorff-Nielsen [3, 4, 5] also considered using a variation on this approximation, of the same form as the univariate expansion of Lugannani and Rice [10]. The drawback of these approximations is that the calculation their calculation requires the calculation of an exact or approximate ancillary, and in some situations it is hard or impossible to construct this ancillary. For the other approximations that we will study in the following, no such ancillary needs to be specified, and hence the approximations are easier to apply in practice.

Severini [12] proposed an approximation $\hat{R}^*$ to Barndorff-Nielsen's $R^*$ based on empirical covariances. Under some assumptions and model regularity properties, $\hat{R}^*$ is distributed according to a standard normal distribution, with error $O(n^{-1})$, conditionally on the observed value of an ancillary statistic $\mathbf{A}$. However, the construction of this $\hat{R}^*$ does not require the specification of $\mathbf{A}$.

DiCiccio and Martin [8] proposed an alternative quantity to $R^*$, denoted by $R^+$, that is available without specification of $\mathbf{A}$. The derivation of $R^+$ involves the Bayesian approach to constructing confidence limits considered by Welch and Peers [15] and Peers [11]. In the presence of nuisance parameters, Peers [11] chose a prior density for the parameters to satisfy a partial differential equation. With this prior, the standard normal approximation to the conditional distribution of $R^+$ has error of order $O(n^{-1})$. If the parameter of interest and the nuisance parameter vector are orthogonal, solving the partial differential equation is relatively easier. In some cases that the parameters are not orthogonal, solving that equation numerically is problematic. Parameter orthogonality will be reviewed in the following section.

For a parameter of interest that is orthogonal to the nuisance parameter vector, Cox and Reid [6] defined the signed root of the conditional likelihood ratio statistic $\overline{R}$. The standard normal approximation to the distribution of $\overline{R}$ has error of order $O(n^{-1/2})$. DiCiccio and Martin [8] defined $\overline{R}^+$ similar as the $R^+$ mentioned above. The standard normal approximation to the conditional distribution of $\overline{R}^+$ has error of order $O(n^{-1})$. The use of $\overline{R}$ and its modifications is often effective in situations where there are many nuisance parameters. In such cases, the use of $R$ and its modified versions can produce unsatisfactory results; see DiCiccio, Field and Fraser [7] for examples.

The above variants on $R$ have never been systematically compared to each other as a group. This paper provides an accuracy comparison among the modifications stated above. Each of these approximations are used to generate an approximate one-sided $p$-value by approximating $\mathrm{P}[R \geq r]$, for $r$ the observed value of $R$. Approximate two-sided $p$-values are calculated by approximating $2\min(\mathrm{P}[R \geq r], \mathrm{P}[R < r])$. One and two-sided hypotheses tests of size $\alpha$ may be constructed by rejecting the null hypothesis when the $p$-value is less than $\alpha$. Both the Barndorff-Nielson format approximation

$$\Phi\{R + R^{-1}\log(U/R)\} \tag{1}$$

and the Lugannani and Rice format approximation

$$\Phi(R) + \phi(R)(R^{-1} - U^{-1}) \tag{2}$$

were considered in this paper, where the variable $U$ may vary for different modifications. We examined as an example the ratio of means of independent exponentials.

We calculated via simulation the size of tests constructed as above, and then compared the results among different approximations.

## 2. Methodology

We first review several statistics whose marginal distributions are very close to standard normal. Consider continuous variables $X_1, \ldots, X_n$ having joint density function that depends on an unknown parameter $\boldsymbol{\omega} = (\omega_1, \ldots, \omega_d)$. Suppose that $\boldsymbol{\omega} = (\psi, \boldsymbol{\chi})$, where $\psi = \omega_1$ is a scalar parameter of interest and $\boldsymbol{\chi} = (\omega_2, \ldots, \omega_d)$ is a nuisance parameter vector. Let $\hat{\boldsymbol{\omega}} = (\hat{\psi}, \hat{\boldsymbol{\chi}})$ be the maximum likelihood estimator of $\boldsymbol{\omega}$, and for fixed $\psi$, let $\hat{\boldsymbol{\chi}}_\psi$ be the constrained maximum likelihood estimator of $\boldsymbol{\chi}$. The signed root of the likelihood ratio statistic is $R = \operatorname{sgn}(\hat{\psi} - \psi_0)\{2(l(\hat{\boldsymbol{\omega}}) - l(\psi_0, \hat{\boldsymbol{\chi}}_0))\}^{1/2}$, where $\hat{\boldsymbol{\chi}}_0$ will be shorthand for $\hat{\boldsymbol{\chi}}_{\psi_0}$ and $l(\boldsymbol{\omega})$ is the log-likelihood function for $\boldsymbol{\omega}$. The standard normal approximation to the distribution of $R$ typically has error of order $O(n^{-1/2})$, and $R$ can be used to construct approximate confidence limits for $\psi$ having coverage error of that order.

The earliest general conditional saddlepoint tail probability approximation was provided by Skovgaard [13], who applied double saddlepoint techniques to the problem of approximating tail probabilities for conditional distributions when the data arise from a full exponential family. In this case the double saddlepoint distribution function approximation can be expressed in terms of the quantities in the joint density function. Skovgaard's double saddlepoint approximation to the conditional distribution function is of form (2), with $U$ a Wald statistic. In this paper, we consider only models more complicated than canonical exponential families, and so won't apply this approximation.

### 2.1. Barndorff-Nielsen's modification

The modified signed root of the likelihood ratio statistic $R^*$ was first proposed by Barndorff-Nielsen [2] and given by

$$R^* = R + R^{-1}\log(U/R),$$

where

$$(3) \qquad U = \frac{|l_{\boldsymbol{\chi};\hat{\boldsymbol{\omega}}}(\hat{\boldsymbol{\omega}}_\psi)\ l_{;\hat{\boldsymbol{\omega}}}(\hat{\boldsymbol{\omega}}) - l_{;\hat{\boldsymbol{\omega}}}(\hat{\boldsymbol{\omega}}_\psi)|}{|j_{\boldsymbol{\chi}\boldsymbol{\chi}}(\hat{\boldsymbol{\omega}}_\psi)|^{\frac{1}{2}}|j(\hat{\boldsymbol{\omega}})|^{\frac{1}{2}}},$$

and $j_{\boldsymbol{\chi}\boldsymbol{\chi}}(\hat{\boldsymbol{\omega}}_\psi) = -l_{\boldsymbol{\chi}\boldsymbol{\chi}}(\psi_0, \hat{\boldsymbol{\chi}}_0)$ and $j(\hat{\boldsymbol{\omega}}) = -l_{\boldsymbol{\omega}\boldsymbol{\omega}}(\hat{\boldsymbol{\omega}})$, with $l_{\boldsymbol{\omega}\boldsymbol{\omega}}(\boldsymbol{\omega})$ the matrix of second-order partial derivatives of $l(\boldsymbol{\omega}; \hat{\boldsymbol{\omega}}, \mathbf{A})$ taken with respect to $\boldsymbol{\omega}$ and $l_{\boldsymbol{\chi}\boldsymbol{\chi}}(\boldsymbol{\omega})$ the submatrix of $l_{\boldsymbol{\omega}\boldsymbol{\omega}}(\boldsymbol{\omega})$ corresponding to $\boldsymbol{\chi}$. Here $U$ represents an approximate conditional score statistic, which, in the multivariate normal case would exactly coincide with $R$. Outside the multivariate normal case, it measures the difference between $R$ and $U$ is a measure of departure from normality. The quantity $l_{;\hat{\boldsymbol{\omega}}}(\boldsymbol{\omega})$ is the $d \times 1$ vector of partial derivatives of $l(\boldsymbol{\omega}; \hat{\boldsymbol{\omega}}, \mathbf{A})$ taken with respect to $\hat{\boldsymbol{\omega}}$, and $l_{\boldsymbol{\chi};\hat{\boldsymbol{\omega}}}(\boldsymbol{\omega})$ is a $d \times (d-1)$ matrix of mixed second-order partial derivatives of $l(\psi, \boldsymbol{\chi}; \hat{\boldsymbol{\omega}}, \mathbf{A})$ taken with respect to $\boldsymbol{\chi}$ and $\hat{\boldsymbol{\omega}}$. The sign of $U$ is the same as that of $R$ and the resulting $U$ is of the form $U = R + O_p(n^{-1/2})$. The relative error in the standard normal approximation to the conditional distribution of $R^*$ is of order $O(n^{-3/2})$. The conditioning is on an exact or approximate ancillary statistic $\mathbf{A}$. The variable $U$ is parameterization invariant and does not depend on $\boldsymbol{\chi}$.

The value of $\psi_0$ satisfying $\Phi(R^*) = \alpha$ is an approximate upper $1-\alpha$ confidence limit which has relative coverage error of order $O(n^{-3/2})$ both conditionally and unconditionally. Barndorff-Nielsen [3, 4, 5] also considered using the alternative to $\Phi(R^*)$ provided by the Lugannani and Rice format approximation (2).

Consider the exponential family model for a random vector $\mathbf{T}$ whose density evaluated at $\mathbf{t}$ is $f_\mathbf{T}(\mathbf{t};\boldsymbol{\theta}) = \exp(\boldsymbol{\theta}^\top \mathbf{t} - \mathcal{H}_\mathbf{T}(\boldsymbol{\theta}) - \mathcal{G}(\mathbf{t}))$. The random vector $\mathbf{T}$ is the sufficient statistic and set $\tau(\boldsymbol{\theta}) = E_{\boldsymbol{\theta}}[\mathbf{T}]$. In the presence of nuisance parameters, the calculation of $U$ requires the specification of the ancillary $\mathbf{A}$. Barndorff-Nielsen [1] suggested an approximate ancillary statistic for use in conditional inference. Kolassa [9], in Chapter 8.4, presented this approximate ancillary $\mathbf{A}$ as

$$\mathbf{B}(\hat{\psi})(\mathbf{T} - \boldsymbol{\tau}(\hat{\psi}, \boldsymbol{\chi}))^\top,$$

with $\boldsymbol{\chi}$ held fixed, and

$$\mathbf{B}(\psi) = [(\partial\boldsymbol{\tau}/\partial\psi)^\perp \boldsymbol{\Sigma}(\partial\boldsymbol{\tau}/\partial\psi)^{\perp\top}]^{-\frac{1}{2}}(\partial\boldsymbol{\tau}/\partial\psi)^\perp.$$

Suppose that $\boldsymbol{\theta}$ is scalar. Let $\tilde{l}(\theta;\hat{\theta},\mathbf{a}) = l(\theta;\hat{\theta},\mathbf{a})/n$. Then

(4) $\quad F_{\hat{\Theta}|\mathbf{A}}(\hat{\theta}|\mathbf{a};\theta) = [\Phi(\sqrt{n}\hat{\omega}) + \phi(\sqrt{n}\hat{\omega})[1/\hat{\omega} - 1/\check{z}]/\sqrt{n}\,][1 + O_p(n^{-1})],$

with $\check{z} = [\tilde{l}^{;1}(\hat{\theta};\hat{\theta},\mathbf{a}) - \tilde{l}^{;1}(\theta;\hat{\theta},\mathbf{a})]/\sqrt{j(\hat{\theta})}$, and the superscripts ;1 on $\tilde{l}^{;1}$ represent differentiation of the likelihood with respect to $\hat{\theta}$, after reexpressing $\mathbf{t}$ in terms of $\hat{\theta}$ and $\mathbf{a}$. Here $\mathbf{a}$ is the observed value of $\mathbf{A}$; $F_{\hat{\Theta}|\mathbf{A}}(\hat{\theta}|\mathbf{a};\theta)$ is the conditional cumulative distribution function and $\Phi(\cdot)$ is the standard normal cumulative distribution function.

In the computation of Barndorff-Nielsen's $R^*$, the calculation of $U$ requires the ancillary $\mathbf{A}$ to be specified, which may present difficulties in practice. In the following, we will introduce several modifications that do not require the specification of $\mathbf{A}$.

### 2.2. An empirical adjustment

Severini [12] proposed approximation $\hat{R}^*$ to Barndorff-Nielsen's $R^*$ based on empirical covariances. Recalling the formula of $U$ (3), the key step is to approximate $l_{\boldsymbol{\chi};\hat{\boldsymbol{\omega}}}(\hat{\boldsymbol{\omega}}_\psi)$ and $l_{;\hat{\boldsymbol{\omega}}}(\hat{\boldsymbol{\omega}}) - l_{;\hat{\boldsymbol{\omega}}}(\hat{\boldsymbol{\omega}}_\psi)$.

Let $l^{(j)}(\boldsymbol{\omega})$ denote the log-likelihood function based on observation $j$ alone. Denote

$$\hat{Q}(\boldsymbol{\omega};\boldsymbol{\omega}_0) = \sum l^{(j)}(\boldsymbol{\omega}) l^{(j)}_{\boldsymbol{\omega}}(\boldsymbol{\omega}_0)^T, \ \hat{I}(\boldsymbol{\omega};\boldsymbol{\omega}_0) = \sum l^{(j)}_{\boldsymbol{\omega}}(\boldsymbol{\omega}) l^{(j)}_{\boldsymbol{\omega}}(\boldsymbol{\omega}_0)^T,$$

and $\hat{i} = \hat{I}(\hat{\boldsymbol{\omega}};\hat{\boldsymbol{\omega}})$. The quantity $\boldsymbol{\omega}_0$ is any point in the parameter space. Then $l_{;\hat{\boldsymbol{\omega}}}(\hat{\boldsymbol{\omega}}) - l_{;\hat{\boldsymbol{\omega}}}(\boldsymbol{\omega})$ and $l_{\boldsymbol{\omega};\hat{\boldsymbol{\omega}}}(\boldsymbol{\omega})$ may be approximated by

$$hatl_{;\hat{\boldsymbol{\omega}}}(\hat{\boldsymbol{\omega}}) - \hat{l}_{;\hat{\boldsymbol{\omega}}}(\boldsymbol{\omega}) = \{\hat{Q}(\hat{\boldsymbol{\omega}};\hat{\boldsymbol{\omega}}) - \hat{Q}(\boldsymbol{\omega};\hat{\boldsymbol{\omega}})\}\hat{i}(\hat{\boldsymbol{\omega}})^{-1}\hat{j}$$

and

$$\hat{l}_{\boldsymbol{\omega};\hat{\boldsymbol{\omega}}}(\boldsymbol{\omega}) = \hat{I}(\boldsymbol{\omega};\hat{\boldsymbol{\omega}})\hat{i}(\hat{\boldsymbol{\omega}})^{-1}\hat{j},$$

where $\hat{j} = j(\hat{\boldsymbol{\omega}}) = -l_{\boldsymbol{\omega}\boldsymbol{\omega}}(\hat{\boldsymbol{\omega}})$.

Denote by $\hat{U}$ the approximation to the statistic $U$ based on the above quantities, and then denote
$$\hat{R}^* = R + R^{-1}\log(\hat{U}/R).$$

The quantity $\hat{R}^*$ can be used in approximation (1). This represents a correction similar to that of (3), with expectations of quantities replaced by sample means. Under some assumptions plus model regularity properties, $\hat{R}^*$ is distributed according to a standard normal distribution, with error $O(n^{-1})$, conditionally on $\mathbf{a}$, the observed value of the ancillary $\mathbf{A}$. However, the construction of $\hat{R}^*$ does not require the specification of $\mathbf{A}$. Again, the alternative approximation (2) is also available as $\Phi(R) + \phi(R)(R^{-1} - \hat{U}^{-1})$.

### 2.3. DiCiccio and Martin's modification

DiCiccio and Martin [8] proposed an alternative variable to $U$, denoted by $T$, which is available without specification of the ancillary $\mathbf{A}$. The modification for approximation (1) is

(5) $$R^+ = R + R^{-1}\log(T/R),$$

where $T$ is defined in (7). As with (3), the final term in $R^+$ represents the departure from normality; unlike (3), this measure represents the departure of the posterior for $\psi$ from normality, and involves the prior distribution. Once again, one might use the alternative probability approximation (2) with $T$ substituting the place of $U$. The replacement of $T$ avoids the necessity of specifying $\mathbf{A}$ in calculating $U$ and hence simplifies the calculations. The derivation of $T$ involves the Bayesian approach to constructing confidence limits considered by Welch and Peers [15] and Peers [11]. When $\omega = \psi$, that is, when the entire parameter is scalar and there are no nuisance parameters, Welch and Peers [15] showed that the appropriate choice is $\pi(\omega) \propto \{i(\omega)\}^{1/2}$, where $i(\omega) = \mathrm{E}\{-\mathrm{d}^2 l(\omega)/\mathrm{d}\omega^2\}$. In the presence of nuisance parameters, Peers [11] showed that $\pi(\boldsymbol{\omega})$ must be chosen to satisfy the partial differential equation

(6) $$\sum_{j=1}^{d} i^{1j}(i^{11})^{-1/2}\frac{\partial}{\partial\omega^j}(\log\pi) + \sum_{j=1}^{d}\frac{\partial}{\partial\omega^j}\{i^{1j}(i^{11})^{-1/2}\} = 0,$$

where $i_{jk}(\boldsymbol{\omega}) = \mathrm{E}\{-\partial^2 l(\boldsymbol{\omega})/\partial\omega^j\partial\omega^k\}$ and $(i^{jk})$ is the $d\times d$ matrix inverse of $(i_{jk})$. The variable $T$ is defined as

(7) $$T = l_\psi(\psi_0,\hat{\boldsymbol{\chi}}_0)\frac{|-l_{\boldsymbol{\chi\chi}}(\psi_0,\hat{\boldsymbol{\chi}}_0)|^{1/2}\pi(\hat{\boldsymbol{\omega}})}{|-l_{\boldsymbol{\omega\omega}}(\hat{\boldsymbol{\omega}})|^{1/2}\pi(\psi_0,\hat{\boldsymbol{\chi}}_0)}.$$

Here $l_\psi(\boldsymbol{\omega}) = \partial l(\boldsymbol{\omega})/\partial\psi$, and $\pi(\boldsymbol{\omega})$ is a proper prior density for $\boldsymbol{\omega} = (\psi, \boldsymbol{\chi})$ which satisfies the equation (6). Then the resulting approximation (2) is

$$\mathrm{P}(\psi \geq \psi_0|X) = \Phi(R) + (R^{-1} - T^{-1})\phi(R) + O(n^{-3/2}),$$

where $T = U + O_p(n^{-1})$, and thus the approximation (1) to the conditional distribution based on $R + R^{-1}\log(T/R)$ has error of order $O(n^{-1})$. To error of the order $O_p(n^{-1})$, $T$ is parameterization invariant under transformations $\boldsymbol{\omega} \mapsto \{\psi, \tau(\boldsymbol{\omega})\}$.

Parameter orthogonality may make a big difference in solving the partial differential equation (6). Orthogonality is defined with respect to the expected Fisher information matrix. We define $\boldsymbol{\theta}_1$ to be orthogonal to $\boldsymbol{\theta}_2$ if the elements of the information matrix satisfy

$$\text{(8)} \qquad i_{\theta_s \theta_t} = \frac{1}{n} \operatorname{E}\left(\frac{\partial l}{\partial \theta_s} \frac{\partial l}{\partial \theta_t}; \boldsymbol{\theta}\right) = \frac{1}{n} \operatorname{E}\left(-\frac{\partial^2 l}{\partial \theta_s \partial \theta_t}; \boldsymbol{\theta}\right) = 0$$

for $s = 1, \ldots, p_1, t = p_1 + 1, \ldots, p_1 + p_2$, where $\boldsymbol{\theta} = (\boldsymbol{\theta}_1, \boldsymbol{\theta}_2)$; $\boldsymbol{\theta}_1$ and $\boldsymbol{\theta}_2$ are of length $p_1$ and $p_2$ respectively. If equation (8) is to hold for all $\boldsymbol{\theta}$ in the parameter space, then the parameterization is sometimes called globally orthogonal. If (8) holds at only one parameter value $\boldsymbol{\theta}_0$, then the vectors $\boldsymbol{\theta}_1$ and $\boldsymbol{\theta}_2$ are said to be locally orthogonal at $\boldsymbol{\theta}_0$. The most direct statistical interpretation of (8) is that the relevant components of the statistic are uncorrelated.

The definition of orthogonality can be extended to more than two sets of parameters, and in particular $\boldsymbol{\theta}$ is totally orthogonal if the information matrix is diagonal. In general, it is not possible to have total parameter orthogonality at all parameter values, but it is possible to obtain orthogonality of a scalar parameter of interest $\psi$ to a set of nuisance parameters. If the parameter of interest and the nuisance parameter vector are orthogonal, solving the partial differential equation (6) is relatively easier. The equation (6) reduces to

$$\text{(9)} \qquad (i_{\psi\psi})^{-1/2} \frac{\partial}{\partial \psi}(\log \pi) + \frac{\partial}{\partial \psi}(i_{\psi\psi})^{-1/2} = 0,$$

whose solutions are of the form $\pi(\psi, \boldsymbol{\chi}) \propto \{i_{\psi\psi}(\psi, \boldsymbol{\chi})\}^{1/2} g(\boldsymbol{\chi})$ (Tibshirani [14]), where $g(\boldsymbol{\chi})$ is arbitrary and the suggestive notation $i_{\psi\psi}(\psi, \boldsymbol{\chi})$ is used in place of $i_{11}(\psi, \boldsymbol{\chi})$. In some cases in which the parameters are not orthogonal, solving equation (6) numerically is problematic.

### 2.4. Conditional likelihood ratio statistic and its modification

For $\psi$ and $\boldsymbol{\chi}$ orthogonal, Cox and Reid [6] defined the conditional likelihood ratio statistic for testing $\psi = \psi_0$ as $\overline{W} = 2\{\bar{l}(\bar{\psi}) - \bar{l}(\psi_0)\}$, where

$$\bar{l}(\psi) = l(\psi, \boldsymbol{\chi}_\psi) - \frac{1}{2} \log |-l_{\boldsymbol{\chi}\boldsymbol{\chi}}(\psi, \hat{\boldsymbol{\chi}}_\psi)|$$

and $\bar{\psi}$ is the point at which the function $\bar{l}(\psi)$ is maximized. The signed root of the conditional likelihood ratio statistic is $\overline{R} = \operatorname{sgn}(\bar{\psi} - \psi_0)\overline{W}^{1/2}$, and the standard normal approximation to the distribution of $\overline{R}$ has error of order $O(n^{-1/2})$. Let $\overline{R}^+ = \overline{R} + \overline{R}^{-1} \log(\overline{T}/\overline{R})$. One may use approximations (1) and (2), say, $\Phi(\overline{R}^+)$ or $\Phi(\overline{R}) + \phi(\overline{R})(\overline{R}^{-1} - \overline{T}^{-1})$, where

$$\overline{T} = \bar{l}^{(1)}(\psi_0)\{-\bar{l}^{(2)}(\bar{\psi})\}^{-1/2} \frac{\pi(\bar{\psi}, \lambda_{\bar{\psi}})}{\pi(\psi_0, \lambda_0)},$$

and $\bar{l}^{(j)} = \operatorname{d}^j \bar{l}(\psi)/\operatorname{d}\psi^j$, $j = 1, 2$. Those approximations have errors of order $O(n^{-1})$.

The use of $\overline{R}$ and its modifications is often effective in situations where there are many nuisance parameters. In such cases, the use of $R$ and its modified versions can produce unsatisfactory results; see DiCiccio, Field and Fraser [7] for examples.

## 3. Example: Exponential samples with orthogonal interest and nuisance parameters

Let $X$ and $Y$ be exponential random variables with means $\mu$ and $\nu$ respectively; the ratio of the means $\nu/\mu$ is the parameter of interest. The parameter transformation $\left\{\mu \to \frac{\lambda}{\sqrt{\psi}},\ \nu \to \lambda\sqrt{\psi}\right\}$ makes the two new parameters $\psi$ and $\lambda$ orthogonal. Then $X$ and $Y$ have expectations $\lambda\psi^{-\frac{1}{2}}$ and $\lambda\psi^{\frac{1}{2}}$, respectively.

Suppose we have $n$ independent replications of $(X,Y)$. Denote $\boldsymbol{\omega} = (\psi, \lambda)$. We can obtain the log-likelihood function as

$$l(\boldsymbol{\omega}) = -n\left[\frac{\psi\bar{x} + \bar{y}}{\lambda\sqrt{\psi}} + 2\log\lambda\right].$$

Each of the approximations in section 2 may be used to generate an approximate one-sided $p$-value by approximating $P[R \geq r]$, for $r$ the observed value of $R$. Approximate two-sided $p$-values may be calculated by approximating $2\min(P[R \geq r], P[R < r])$. One and two-sided hypotheses tests of size $\alpha$ may be constructed by rejecting the null hypothesis when the $p$-value is less than $\alpha$. Both the Barndorff-Nielson format approximation (1) and the Lugannani and Rice format approximation (2) were considered. We calculated via simulation the size of tests constructed as above, and compared the results among different approximations.

Some of the approximations in section 2 require specific algebraic calculations. We present the related calculations below. Other applications are generic, and no specific algebraic calculations are needed.

### 3.1. Some algebraic calculations

**Barndorff-Nielson's modification**

The expectations of the sufficient statistics $\mathbf{T} = (\overline{X}, \overline{Y})$ in the new parameterization are $\boldsymbol{\tau}(\psi, \lambda) = \{\lambda/\sqrt{\psi}, \lambda\sqrt{\psi}\}$, and

$$\mathrm{d}\,\boldsymbol{\tau}(\psi, \lambda)/\,\mathrm{d}\,\psi = \{-\lambda/(2\psi^{\frac{3}{2}}), \lambda/(2\psi^{\frac{1}{2}})\}.$$

A vector perpendicular to this is $(\mathrm{d}\,\boldsymbol{\tau}(\psi, \lambda)/\,\mathrm{d}\,\psi)^{\perp} = \{\psi, 1\}$. The variance of the sample mean vector is

$$\boldsymbol{\Sigma}(\psi, \lambda) = \frac{1}{n}\begin{pmatrix} \lambda^2/\psi & 0 \\ 0 & \lambda^2\psi \end{pmatrix}.$$

In our case, $\mathbf{B}(\psi) = \sqrt{n}\left\{\sqrt{\psi}/(\sqrt{2}\lambda), 1/(\lambda\sqrt{2\psi})\right\}$, and

$$A = \sqrt{2n}\left(\sqrt{\overline{X}\,\overline{Y}}/\lambda - 1\right).$$

Using Barndorff-Nielson's formula [4],

$$\tilde{l}(\psi; \hat{\psi}, a) = -\frac{|a + \sqrt{2n}|(\psi + \hat{\psi})}{\sqrt{2n\psi\hat{\psi}}} - 2\log\lambda,$$

and

$$\hat{\omega} = \mathrm{sign}(\hat{\psi} - \psi)\psi^{1/4}\left|(a + \sqrt{2n})(\psi^{\frac{1}{2}} - \hat{\psi}^{\frac{1}{2}})\right|/(\hat{\psi}^{1/4}n^{1/2}).$$

The negative of the second derivative of the log likelihood is

$$j(\psi) = \left|(a + \sqrt{2n})\right|(3\psi - \hat{\psi})/(4\sqrt{2n\psi\hat{\psi}^5}),$$

and the derivative of $\tilde{l}(\psi; \hat{\psi}, a)$ with respect to $\hat{\psi}$ is

$$\left|(a + \sqrt{2n})\right|(\psi - \hat{\psi})/(2\sqrt{2n\psi\hat{\psi}^3}).$$

Then the quantity $\check{z}$ contributing to the tail probability approximation (4) is $\check{z} = -\sqrt{|a + \sqrt{2n}|}\,(\psi - \hat{\psi})/(2\sqrt{2n\psi\hat{\psi}})$.

**DiCiccio and Martin's modification**

Based on the above, the information matrix is

$$i(\boldsymbol{\omega}) = \mathrm{E}[-l''(\boldsymbol{\omega})] = n\begin{pmatrix}1/2\psi^2 & 0 \\ 0 & 2/\lambda^2\end{pmatrix}.$$

The maximum likelihood estimators are $\hat{\psi} = \overline{Y}/\overline{X}$ and $\hat{\lambda} = (\hat{\psi}\overline{X} + \overline{Y})/(2\hat{\psi}^{1/2}) = \sqrt{\overline{X}\,\overline{Y}}$. For fixed $\psi$, let $\hat{\lambda}_\psi$ be the constrained maximum likelihood of $\lambda$. Here, $\hat{\lambda}_\psi = (\psi\overline{X} + \overline{Y})/(2\psi^{1/2})$. If $\psi = \psi_0 = 1$, then $\hat{\lambda}_{\psi_0} = \hat{\lambda}_0 = (\overline{X} + \overline{Y})/2$.

In this case, the parameters are orthogonal. Using the simplified partial differential equation (9), we chose $g(\lambda) = 1$, and hence $\pi(\psi, \lambda) = \sqrt{n}/(\sqrt{2}\,\psi)$.

In addition to use the prior solved from equation (9), we also studied the outcome from a uniform prior, that is to say, the prior with a constant density, which is obviously not a solution to equation (9).

### 3.2. Simulation results

**Simulation procedure**

For sample size $n = 10$,
(1) Generate 10 draws from the pair of $\{X, Y\}$, where $X$ and $Y$ both follow standard exponential distribution;
(2) Calculate one and two-sided $p$-values for each approximation;
(3) Compare the $p$-values to the $\alpha$ level, say 0.05; denote by $q$ the number of miss coverages; if the $p$-value is less than 0.05, then $q = q + 1$;
(4) Repeat steps (1)-(3) for 10,000 times and report the final value of $q$; let $q^* = (q/10000) * 100$, the Type I error probability in percentage.
(5) Repeat steps (1)-(4) for 100 times and report the average of $q^*$ as the measurement of the accuracy for the modifications.

Approximations (1) and (2) have a removable singularity at $R = 0$. Consequently, these and similar formulae require care when evaluating near $R = 0$. Specifically, we found (1) and (2) to exhibit adequate numerical stability as long as $|R| > 10^{-4}$. Out of 1,000,000 simulated data sets, 60 presented $R$ (or a modification of $R$) closer to zero. In these cases, for all but the most extreme conditioning events, the resulting conditional $p$-value is large enough as to not imply rejection of the null hypothesis, and so these simulated data sets were treated as not implying rejection of the null hypothesis.

258 J. Zhang and J. E. Kolassa

TABLE 1
*Type I error probability (BN)*

| Approximation | One-sided Average | Two-sided Average |
|---|---|---|
| $\Phi(R)$ | 5.241 | 5.168 |
| $\Phi(\overline{R})$ | 4.807 | 4.575 |
| $\Phi(R + R^{-1}\log(U/R))$ | 5.046 | 4.760 |
| $\Phi(R + R^{-1}\log(\hat{U}/R))$ | 5.018 | 4.882 |
| $\Phi(R + R^{-1}\log(T/R))$ | 4.615 | 4.312 |
| $\Phi(R + R^{-1}\log(T_u/R))$ | 11.017 | 6.828 |
| $\Phi(\overline{R} + \overline{R}^{-1}\log(\overline{T}/\overline{R}))$ | 4.883 | 4.411 |
| $\Phi(\overline{R} + \overline{R}^{-1}\log(\overline{T}_u/\overline{R}))$ | 11.723 | 7.249 |

TABLE 2
*Type I error probability (LR)*

| Approximation | One-sided Average | Two-sided Average |
|---|---|---|
| $\Phi(R)$ | 5.241 | 5.168 |
| $\Phi(\overline{R})$ | 4.807 | 4.575 |
| $\Phi(R) + \phi(R)(R^{-1} - U^{-1})$ | 5.046 | 4.760 |
| $\Phi(R) + \phi(R)(R^{-1} - \hat{U}^{-1})$ | 5.017 | 4.881 |
| $\Phi(R) + \phi(R)(R^{-1} - T^{-1})$ | 4.613 | 4.308 |
| $\Phi(R) + \phi(R)(R^{-1} - T_u^{-1})$ | 11.274 | 6.943 |
| $\Phi(\overline{R}) + \phi(\overline{R})(\overline{R}^{-1} - \overline{T}^{-1})$ | 4.878 | 4.403 |
| $\Phi(\overline{R}) + \phi(\overline{R})(\overline{R}^{-1} - \overline{T}_u^{-1})$ | 12.190 | 7.510 |

**Results**

Tables 1 and 2 below report the average of the Type I error probabilities (in percentage) of the 100 rounds simulation. The quantities $T_u$ and $\overline{T}_u$ are assumed with uniform prior densities.

From the tables, we can see that for both the Barndorff-Nielsen format approximation (BN) and the Lugannani and Rice format approximation (LR), the empirical adjustment has best performance. Barndorff-Nielsen's modification has the best asymptotic error rate ($O(n^{-3/2})$ rather than $O(n^{-1})$), and hence we might expect that the best performance from this approximation. Instead we observe the best performance from other modifications with worse asymptotic error. The authors hope to explore this discrepancy in later work.

One may also notice that the performance of DiCiccio and Martin's modification is not as good as expected. One explanation could be that, generally in small sample settings Bayesian method is more sensitive to the choice of the prior density than in the large sample situations. The importance of the choice of prior can be demonstrated by the poor performance of the approximations with the incorrect uniform priors.

**References**

[1] BARNDORFF-NIELSEN, O. E. (1980). Conditionality resolutions. *Biometrika* **67** 293–310. MR0581727
[2] BARNDORFF-NIELSEN, O. E. (1986). Inference on full or partial parameters based on the standardized signed log likelihood ratio. *Biometrika* **73** 307–322. MR0855891
[3] BARNDORFF-NIELSEN, O. E. (1988). Discussion of the paper by N. Reid. *Statist. Sci.* **3** 228–229.